\documentclass[journal,twoside]{IEEEtran}
\usepackage{balance}
\usepackage{cite}
\usepackage{graphicx}

\begin{document}

\title{UWB On-Body Radio Channel Modelling Using Ray Theory and Sub-band FDTD Method}

\author{Yan~Zhao,~\IEEEmembership{Student~Member,~IEEE,}
        Yang~Hao,~\IEEEmembership{Member,~IEEE,}
        Akram~Alomainy,~\IEEEmembership{Student~Member,~IEEE,}
        and~Clive~Parini,~\IEEEmembership{Member,~IEEE}
}

\markboth{IEEE Transactions on Microwave Theory and Techniques, Special Issue on Ultra-Wideband}{Y.~Zhao \MakeLowercase{\textit{et al.}}: UWB On-Body Radio Channel Modelling Using Sub-band FDTD Method}
\maketitle

\begin{abstract}
    This paper presents the ultra-wideband (UWB) on-body radio channel modelling using a sub-band Finite-Difference Time-Domain (FDTD) method and a model combining the uniform geometrical theory of diffraction (UTD) and ray tracing (RT). In the sub-band FDTD model, the frequency band (3 - 9 GHz) is uniformly divided into 12 sub-bands in order to take into account the material frequency dispersion. Each sub-band is simulated separately and then a combination technique is used to recover all simulations at the receiver. In the UTD/RT model, the RT technique is used to find the surface diffracted ray path while the UTD is applied for calculating the received signal. Respective modelling results from two-dimensional (2-D) and three-dimensional (3-D) sub-band FDTD and UTD/RT models indicate that antenna patterns have significant impacts on the on-body radio channel. The effect of different antenna types on on-body radio channels is also investigated through the UTD/RT approach.
\end{abstract}



\section{Introduction}
\PARstart{N}{ext} generation wireless and mobile systems are evolving towards personal and user-centric networks, where constant and reliable connectivity and services are essential. The idea of a number of nodes scattered around the human body and communicating wirelessly with each other sounds appealing and promising to many technologists and developers. This has led to the rapid increase in studies on wireless body area networks (WBAN). One promising application is patient monitoring, where the user is no longer restricted to a specified place which results in faster recovery and less expensive treatment at any time. Other applications of wireless body-centric networks include wearable entertainment systems and high performance mobile personal computers (PCs).

    For low-power, reliable and robust on-body communication systems, a deterministic and generic channel model is required to provide a clearer picture of the on-body radio propagation and its behaviour with regards to different environments and system components. There have been a number of literatures characterising and analysing the on-body channel and also investigating the electromagnetic wave propagation around the body \cite{Hall} - \cite{Nechayev}. However, to the authors' knowledge, UWB on-body radio channels have not been well studied due to the difficulty in characterising frequency dependent electrical properties of human tissues and other effects from antenna types and body movements \textit{etc.}

    The conventional and empirical channel models available for many narrowband and wideband systems are insufficient to describe UWB channel behaviour due to the ultra-wideband nature of the transmitted signals. The ray tracing (RT) technique and FDTD method have been widely studied and applied to indoor/outdoor propagation modelling for narrowband and UWB systems. Sarkar \textit{et al} presented a survey of various propagation models for mobile communications \cite{Sarkar}. Wang \textit{et al} introduced a hybrid technique based on the combination of RT and FDTD methods for narrowband systems \cite{Wang}. Recently, Attiya \textit{et al} proposed a simulation model for UWB indoor radio channels using RT \cite{Attiya}. For UWB on-body radio channel modelling, Fort \textit{et al} simulated pulse propagation around the torso at the frequency range 2 - 6 GHz using Remcom XFDTD \cite{Fort}. However, the variation of UWB on-body channel at different frequencies caused by material dispersion was not taken into account. In this paper, we present a novel deterministic on-body channel model using a sub-band FDTD method. Numerical results obtained are compared with those from a hybrid UTD/RT model. The main advantage of the proposed sub-band FDTD over UTD/RT is its accuracy when modelling complicated on-body radio channels at UWB frequency band. Compared with the dispersive FDTD, although the dispersive FDTD has been developed to model general dispersive materials, the determination of the coefficients for the rational functions to fit measurement data requires further effort such as applying the Pad$\acute{\textrm{e}}$ approximations or the frequency-domain Prony method (FDPM) \cite{Weedon}, \cite{Fan}. The sub-band approach can be directly applied to different human tissues with any type of frequency dependence.

    The rest of the paper is organised as follows: section II introduces the measurement setup for UWB on-body radio channels and results will be used for the evaluation of proposed channel models, section III proposes the sub-band FDTD and UTD/RT model for UWB on-body channels, section IV presents numerical results and their comparison with the measurement and section V draws a conclusion.

\section{Measurement Setup}
    The Fourier Transform relation between time domain and frequency domain signals allows the measurement of channel impulse responses using frequency domain sounding setup. The radio propagation channel measurement in frequency domain has been proven \cite{Howard}, \cite{Hashemi} to be accurate as several time domain techniques if real-time signals are not required and long distances not included.
\begin{figure}[t]
    \centering
    \includegraphics[width=8.2cm,height=6cm]{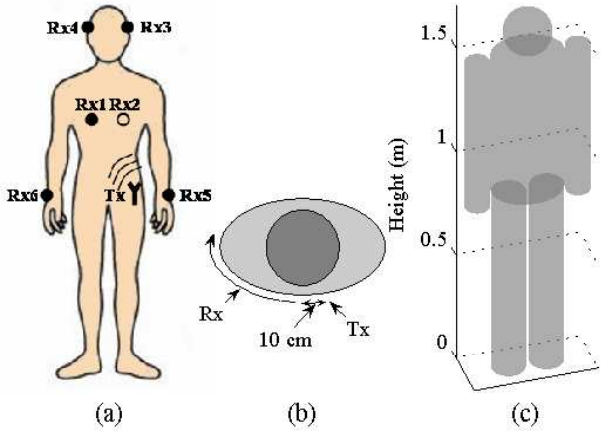}
    \caption{Two-dimensional (2-D) and three-dimensional (3-D) human body models: (a) Antenna positions for on-body radio channel modelling and measurement (b) A 2-D ellipse cylinder used for modelling both transmitter and receiver mounted on the trunk (c) A 3-D human body model used in both sub-band FDTD and UTD/RT models.}
    \label{HumanBody}
\end{figure}

\begin{figure}[t]
    \centering
    \includegraphics[width=8.8cm,height=10.2cm]{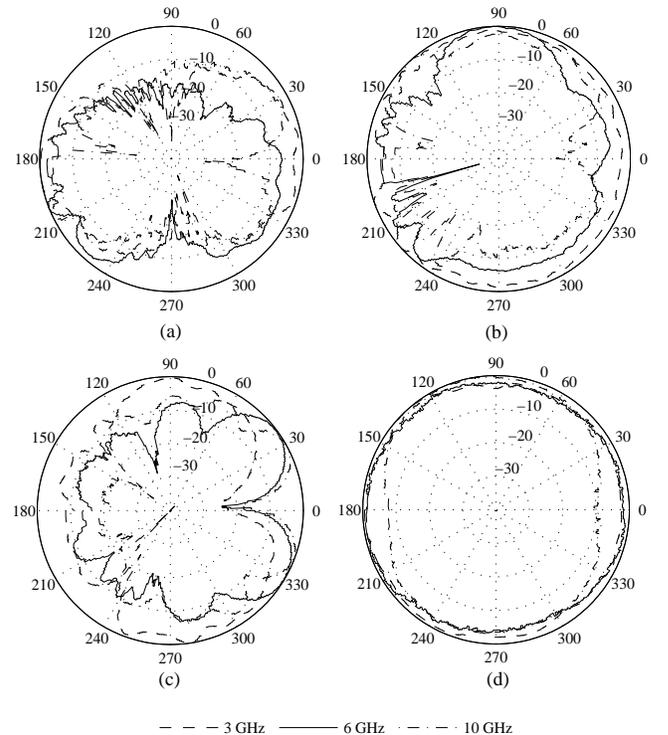}
    \caption{Measured radiation patterns at 3, 6, and 10 GHz for HSCA (a) E-field (b) H-field and PICA (c) E-field (d) H-field. The solid line (6 GHz) is the curve used in UTD/RT model.}
    \label{Patterns}
\end{figure}

    A vector network analyser (VNA) is used to measure the S21 (complex frequency responses) of the on-body UWB channel covering the frequency band 3 - 9 GHz with intervals of 3.75 MHz at a sweeping rate of 800ms over 1601 assigned tones. Measurement settings and procedure are detailed in \cite{Alomainy1}, \cite{Alomainy3}. Various on-body antenna positions and different body postures are applied to obtain a deterministic UWB channel model. Fig. \ref{HumanBody} (a) shows the transmitting and receiving antenna positions as placed during measurement for the characterisation of different on-body links. 710 frequency responses are collected for post measurement analysis and data processing. Two sets of measurements are performed in the anechoic chamber. The UWB antennas used for the on-body measurement campaign are the printed Horn Shaped Self-Complementary Antenna (HSCA) and Planar Inverted Cone Antenna (PICA). The antennas are designed and fabricated following the description outlined in \cite{Alomainy1}, \cite{Alomainy3}. HSCA exhibits approximately constant impedance and absolute gain across the UWB band. The PICA antenna provides outstanding impedance and radiation pattern performance with gain of 0 to 3 dBi \cite{Alomainy1}, \cite{Alomainy3}. Comparing the far field radiation patterns for both azimuth plane and elevation plane, Fig. \ref{Patterns} of the two antennas at different frequencies shows that better radiation bandwidth obtained for the PICA case in comparison to HSCA. During the path loss measurement around the trunk (Fig. \ref{HumanBody} (b)), the printed PICA is used and placed conformal to the body; for the whole body channel measurement (Fig. \ref{HumanBody} (a)), PICA and HSCA are used and placed normal and conformal to the body surface, respectively. The effects of antenna types on the on-body UWB channels are analysed and investigated in details in \cite{Alomainy1}, \cite{Alomainy3}, with modelling aspects discussed intensively in this paper.

\section{Modelling Techniques}
    The main aim of this study is to investigate on-body radio channels and develop appropriate modelling tools. In this paper, the on-body channel is modelled in free space environment and compared with measurements performed in anechoic chamber. A wideband Gaussian monocycle is chosen as the pulse excitation \cite{TimeDomain}: in the sub-band FDTD model, different pulses are used according to the centre frequency of each sub-band; for the UTD/RT, a single pulse of central frequency at 6 GHz is used.

\subsection{The Sub-band FDTD Model}
    In UWB radio channels, the inherent material dispersions represent the changes of permittivity and conductivity \textit{etc.} with frequency. Such dispersions cannot be directly modelled using existing dispersive FDTD based on Debye/Lorentz relations. To apply the sub-band FDTD method, one can follow these steps \cite{Terre}, \cite{Sugahara}:
\begin{enumerate}
\item
    First divide the whole frequency band into several sub-bands, each of which is narrow enough to assume same frequency characteristics.
\item
    Use the conventional FDTD method to obtain the time domain delay profiles for each sub-band.
\item
    Then Fourier transform sub-band delay profiles into the frequency domain; extract the ``accurate'' part and combine them to have a new frequency response.
\item
    Finally transform the frequency responses back into the time domain to have a delay profile that is valid over the entire ultra wideband.
\end{enumerate}

\begin{figure}[t]
    \centering
    \includegraphics[width=8.2cm,height=6.82cm]{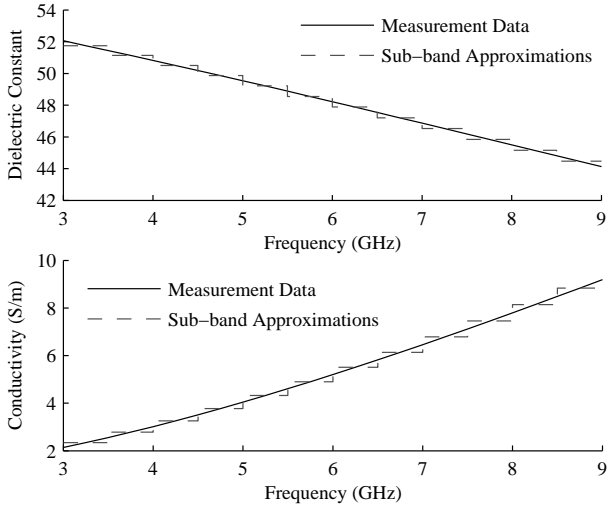}
    \caption{Measured permittivity and conductivity of human muscle in the UWB band and sub-band approximations for the sub-band FDTD model.}
    \label{Permittivity}
\end{figure}

    The choice on the number of sub-bands depends on the accuracy requirement to approximate dispersive material properties. Generally, the greater number of sub-bands, the more accurate the sub-band FDTD model is. However, since each sub-band needs to be simulated separately, as the number of sub-bands increases, further simulation efforts are required. For our on-body propagation analysis, sufficient accuracy can be ensured by dividing the whole frequency band (3 - 9 GHz) into 12 sub-bands with 500 MHz bandwidth for each sub-band. For instance, the relative dielectric constant of human muscle ranges from 52.058 at 3 GHz to 44.126 at 9 GHz \cite{Tissues}. 12 sub-bands are used to match the frequency dispersion curve by assuming the dielectric constant within each sub-band to be constant (obtained at the centre frequency of each sub-band). The overall error from such a curve fitting is less than 1\%. Fig. \ref{Permittivity} shows the frequency dependent dielectric constant and conductivity of human muscle from measurement and their staircasing approximations used in the proposed sub-band FDTD model.

    For step 3, the combination of received signals in the frequency domain can be obtained using (\ref{eq_signal}),
\begin{equation}
    F_{r}(\omega)=\sum^{N}_{i=1}F_{r,i}(\omega)\cdot A_{i}(\omega)
    \label{eq_signal}
\end{equation}
where $F_{r,i}(\omega)$ is the received frequency domain signal at the $i$th sub-band, $A_{i}(\omega)$ is a rectangle window function associated with the bandwidth of the $i$th sub-band and $N$ is the total number of sub-bands. Finally, the combined frequency domain signal is inverse Fourier transformed into the time domain to obtain time delay profile.

    Although the FDTD method has been successfully applied to the antenna design \cite{Tirkas} for various applications, the inclusion of antenna patterns \textit{etc.} in radio propagation analysis using a global FDTD method is not feasible due to the constrains of limited computer resources. In this paper, the antenna is approximated as a point source of narrow Gaussian pulse and it is viable in particular for PICA, which radiates almost omni-directionally across the frequency band.

\subsection{The UTD/Ray Tracing Model}
    Basic RT techniques include two approaches: the image method \cite{Lawton} and the method of shooting and bouncing rays (SBR) \cite{Ling}. In the UTD/RT model proposed in this paper, the SBR method is used. When the on-body channel is modelled using the hybrid UTD/RT approach, the RT is used to find the surface ray path while the UTD is applied for calculating surface diffracted signal strength. In \cite{Ghaddar}, it has been shown that the human body can be modelled as a metallic cylinder regardless of its associated dielectric parameters. In our UTD/RT model, conducting sphere and cylinders are used to represent different parts of the human body thus the UTD surface diffraction coefficients \cite{McNamara} can be used. The UTD solutions for radiation problem in shadow zone are given by (\ref{eq_radiation_shadow1}) and (\ref{eq_radiation_shadow2}),
\begin{eqnarray}
\lefteqn{\textbf{E}^{d}(P_{d})=C_{0}Z\textbf{J}(Q')\cdot\hat{\textbf{n}}(Q')\hat{\textbf{n}}(Q)H(\xi_{c})}\nonumber\\
&&~~~~~~~~~~~~~~~~~~~~~\times\left[\frac{a_{0}(Q)}{a_{0}(Q')}\right]^{1/6}e^{-jkt}\frac{e^{-jks^{d}}}{\sqrt{s^{d}}}
    \label{eq_radiation_shadow1}
\end{eqnarray}
for electric current source, and
\begin{eqnarray}
\lefteqn{\textbf{E}^{d}(P_{d})=C_{0}\textbf{M}(Q')\cdot\left[\hat{\textbf{b}}\hat{\textbf{n}}(Q)H(\xi_{c})+\hat{\textbf{t}}(Q')\hat{\textbf{b}}S(\xi_{c})\right]}\nonumber\\
&&~~~~~~~~~~~~~~~~~~~~~\times\left[\frac{a_{0}(Q)}{a_{0}(Q')}\right]^{1/6}e^{-jkt}\frac{e^{-jks^{d}}}{\sqrt{s^{d}}}
    \label{eq_radiation_shadow2}
\end{eqnarray}
for magnetic current source where $H^{l}(\xi)$ and $S^{l}(\xi)$ are defined in terms of the hard and soft Fock radiation functions. The UTD solutions for surface coupling problem are given by (\ref{eq_coupling_E}) and (\ref{eq_coupling_H}),
\begin{equation}
\textbf{E}(Q)=C_{0}Z\textbf{J}(Q')\cdot\left[\hat{\textbf{n}}(Q')\hat{\textbf{n}}(Q)\right]F_{s}(\xi_{c})e^{-jkt}
    \label{eq_coupling_E}
\end{equation}
for electric current source, and
\begin{equation}
\textbf{H}(Q)=C_{0}Y\textbf{M}\cdot\left[\hat{\textbf{b}}(Q')\hat{\textbf{b}}(Q)F_{s}(\xi_{c})+\hat{\textbf{t}}(Q')\hat{\textbf{t}}(Q)G_{s}(\xi_{c})\right]e^{-jkt}
    \label{eq_coupling_H}
\end{equation}
for magnetic current source where $F_{s}$ applies to the TE coupling configuration and $G_{s}$ is applicable to the TM situation. Other parameters and geometrical arrangements for calculating surface diffracted field in (\ref{eq_radiation_shadow1}) - (\ref{eq_coupling_H}) are defined in \cite{McNamara}. Note that before tracing each ray, the associated field is first separated into a TE and a TM part, then the surface diffracted field can be calculated in terms of each part and combined at the receiver.

\begin{figure}[t]
    \centering
    \includegraphics[width=8.2cm,height=2.43cm]{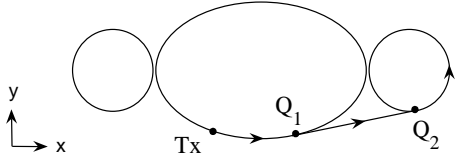}
    \caption{2-D view of surface diffracted ray path with the locations of the transmitter, Tx, the shedding point, $\textrm{Q}_{1}$ and the attachment point, $\textrm{Q}_{2}$.}
    \label{RTPROC}
\end{figure}

    For a simple 2-D case as shown in Fig. \ref{HumanBody} (b) with given geometrical parameters and field polarisation information, the UTD diffraction coefficient for convex surface coupling problem can be directly used to calculate diffracted field without the need for tracing rays at different directions. For the 3-D scenario as shown in Fig. \ref{HumanBody} (c), since human body is modelled in free space, only the rays in the tangential plane at the transmitter need to be traced. Provided that the tracing plane does not intersect with other body parts besides the trunk, all the traced rays start with creeping waves. The procedure for finding the surface diffracted ray path can be summarised as follows:
\begin{enumerate}
\item
    First establish a simplified 3-D human body model (Fig. \ref{HumanBody} (c)) with all the required geometrical dimensions. Assign the locations of transmitter and receiver.
\item
    Generate the first ray at the initial tracing angle ($0^{\circ}$ to the horizontal plane) from the transmitter.
\item
    Along the current ray direction, search for possible surface ray path (up to one surface diffraction) to the receiver according to the generalised Fermat principle \cite{McNamara}; search for possible reflected ray path (up to one reflection) to the receiver from other body parts.
\item
    If any of the ray paths is present, use UTD surface diffraction/reflection coefficient to calculate the diffracted/reflected signal strength.
\item
    Search for another surface ray path on different body parts along the ray direction as shown in Fig. \ref{RTPROC}, and repeat step 3 until the associated field strength falls below a pre-specified threshold.
\item
    Generate a new ray at a different angle (\textit{e.g.} $ 0.5^{\circ}$ increasement) and start over from step 3.
\end{enumerate}

    Note that from step 3, multiple reflected rays are not taken into account in our UTD/RT model due to the complexity in finding multiple reflected ray path for the on-body scenario. In step 5, in order to find the surface ray path, the shedding point $\textrm{Q}_{1}$ and attachment point $\textrm{Q}_{2}$ are first calculated geometrically in a 2-D ($x$-$y$) plane (projection of 3-D) as illustrated in Fig. \ref{RTPROC}. Then the position of these points in $z$-direction can be calculated from the direction of the ray. At the receiver, each of the received rays contains the effect of multiple surface diffractions/single reflection from different body parts. The polarisation information for calculating diffracted field is obtained from the orientation of antennas placed in measurement.

    The received frequency domain signal can be calculated using (\ref{eq_signal_rt}) \cite{McNamara},
\begin{equation}
E(\omega)=\sum^{N}_{i=1} E_{0}(\omega)G_{ti}(\omega)G_{ri}(\omega)A_{i}\prod_{j}D_{j}(\omega)\prod_{l}R_{l}(\omega)e^{-j\frac{\omega}{c}d_{i}}
    \label{eq_signal_rt}
\end{equation}
where $E_{0}(\omega)$ is the transmitted frequency domain signal, $G_{ti}(\omega)$ and $G_{ri}(\omega)$ are the transmitting and receiving antenna field radiation patterns in the direction of the $i$th ray, $A_{i}$ is a distance factor, $D(\omega)$ and $R(\omega)$ are the diffraction and reflection coefficients, $j$ and $l$ depend on the number of diffractions and reflections before reaching the receiver, respectively, $e^{-j\frac{\omega}{c}d_{i}}$ is the propagation phase factor due to the path length $d_{i}$, $c$ is the speed of wave propagation and $N$ is the total number of received rays. Note that in our simulations, the frequency dependency of $G_{ti}(\omega)$ and $G_{ri}(\omega)$ in (\ref{eq_signal_rt}) is not taken into account and the patterns at 6 GHz for both antennas are used. Therefore, the frequency domain behaviour caused by the radiation process of UWB antennas \cite{Siwiak} is not included in our analysis.

\section{Numerical Results and Analysis}
    The sub-band FDTD method is validated by comparing the reflection coefficient calculated from 1-D sub-band FDTD simulation and analytical equations. The analytical equation is given by \cite{Sato},
\begin{equation}
    R(\omega)=\frac{1-e^{-j2\delta}}{1-R'^{2}(\omega)e^{-j2\delta}}R'(\omega)
    \label{eq_reflection}
\end{equation}
where $\delta=\frac{2\pi d}{\lambda}\sqrt{n^{2}(\omega)-\sin^{2}(\theta)}$, $\lambda$ is the wavelength in free space, $d$ is the thickness of the dielectric plate, $n(\omega)$ is its complex and dispersive refractive index and $\theta$ is the angle of incidence. In (\ref{eq_reflection}), $R'(\omega)$ is given by,
\begin{equation}
R'_{s}(\omega)=\frac{\cos(\theta)-\sqrt{n^{2}(\omega)-\sin^{2}(\theta)}}{\cos(\theta)+\sqrt{n^{2}(\omega)-\sin^{2}(\theta)}}
    \label{eq_reflection1}
\end{equation}
or
\begin{equation}
R'_{p}(\omega)=\frac{n^{2}(\omega)\cos(\theta)-\sqrt{n^{2}(\omega)-\sin^{2}(\theta)}}{n^{2}(\omega)\cos(\theta)+\sqrt{n^{2}(\omega)-\sin^{2}(\theta)}}
    \label{eq_reflection2}
\end{equation}
where $R'_{s}(\omega)$ and $R'_{p}(\omega)$ are the Fresnel's reflection coefficients for the interface between air and a dielectric media whose frequency dependent complex refractive index is $n(\omega)$ when the electric field is perpendicular and parallel to the incident plane, respectively.
\begin{figure}[t]
    \centering
    \includegraphics[width=6.41cm,height=2.06cm]{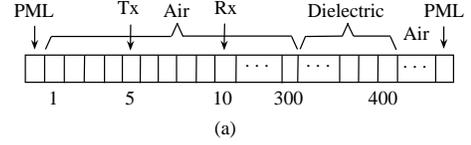}
    \includegraphics[width=8.2cm,height=6cm]{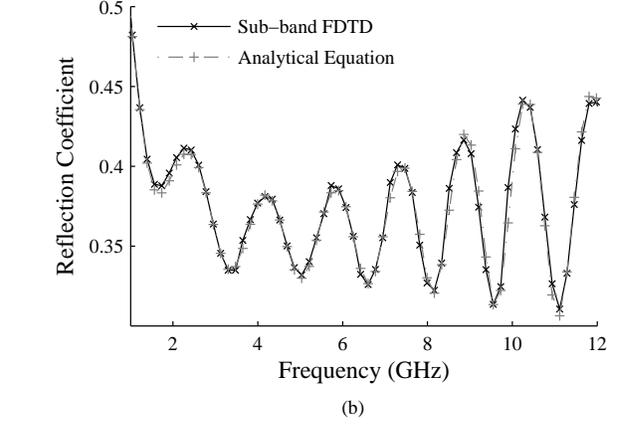}
    \caption{(a) A 1-D sub-band FDTD simulation domain (b) Comparison of reflection coefficient for normal incidence at an air - brick interface calculated from the sub-band FDTD method and analytical equation.}
    \label{Reflection}
\end{figure}

    For the 1-D sub-band FDTD simulation, 300 cells are used to model the free space and 100 cells for the dielectric slab (modelled as bricks with thickness 0.20 m) as shown in Fig \ref{Reflection} (a). The cell size is $2.0\times10^{-3}$ m and the time step is $3.3\times10^{-12}$ s. The reflected signal is obtained by subtracting the direct signal from the total received signal and the wideband reflection coefficient is calculated by dividing the reflected and incident field strength in the frequency domain. It can be seen in Fig. \ref{Reflection} (b), the comparison shows good agreement and validates the sub-band FDTD method.

    A detailed comparison between different modelling techniques for indoor UWB radio propagation has been presented in \cite{Zhao}. When these techniques are applied to the UWB on-body communications, a simplified human body model is used with only muscles are considered. The frequency dependent dielectric constant and conductivity of human muscle can be found from measurement \cite{Tissues} and shown in Fig. \ref{Permittivity}.

\subsection{Two-Dimensional On-Body Propagation Channels}
    Applying both sub-band FDTD and UTD/RT methods to a simple (2-D) scenario is first considered. Both the transmitter and receiver are mounted on the trunk. As the receiver moves along the trunk in the same horizontal plane, the scenario can be treated as a 2-D case. As shown in Fig. \ref{HumanBody} (b), the human body (trunk) is modelled as a 2-D ellipse cylinder with semi-major axis 0.15 m and semi-minor axis 0.12 m according to the dimensions of a human candidate volunteered in the measurement. Both transmitter and receiver are placed on the `trunk' and the transmitter is 10 cm offset from the centre. During the measurement, the receiver is always kept on the `trunk' while moving along the route as shown in Fig. \ref{HumanBody} (b). The transverse magnetic (TM) polarised field is considered for both sub-band FDTD and UTD/RT models due to the orientation and the radiated field of the antenna used in measurement. The antenna pattern contribution is excluded in this analysis because for any receiver location, the received signal only contains the contributions from two creeping waves travelling from the transmitter at opposite directions tangential to the ellipse's surface. While mutual coupling between transmitting and receiving antennas is considered \cite{McNamara}: with given transmitter/receiver locations and geometrical dimensions (ellipse cylinder), the UTD diffraction coefficient for TM-polarised field can be directly used to calculate diffracted signal strength. The approximate elliptic `trunk' is also modelled using the sub-band FDTD with the cell size of $3.0\times10^{-3}$ m. The number of cells in the computational region is $140\times160$, which is truncated by a 10-cell Berenger's perfect matched layer (PML) \cite{Berenger}.

    On-body path loss is calculated at different receiver locations using (\ref{eq_pl}), \begin{equation}
    PL(d)=-10\log_{10}(\frac{E_{r}}{E_{t}})
    \label{eq_pl}
\end{equation}
where $E_{t}$ and $E_{t}$ are total transmitted and total received signal energy respectively.

\begin{figure}[t]
    \centering
    \includegraphics[width=8.2cm,height=6cm]{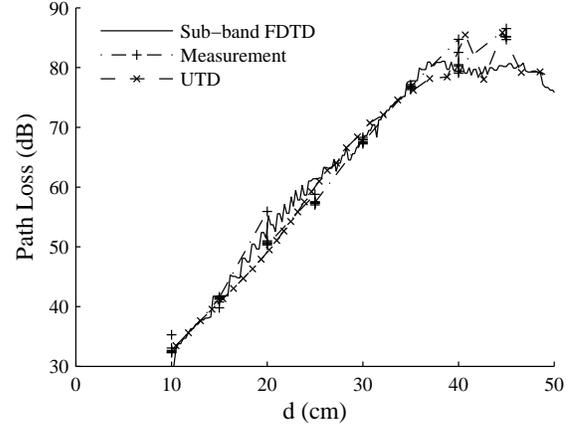}
    \caption{Comparison of path losses along the trunk (Fig. \ref{HumanBody} (b)) from the sub-band FDTD model, UTD model and measurement.}
    \label{2DPL}
\end{figure}

    Fig. \ref{2DPL} shows the path loss results along the trunk (Fig. \ref{HumanBody} (b)) from the sub-band FDTD model, UTD/RT model and measurement. Good agreement is achieved when the creeping distance of the transmitter and receiver is small. However, when the distance approaches the maximum, ripples are observed from UTD/RT and measurement, which are caused by the adding up or cancelling of two creeping rays travelling along both sides of the elliptical `trunk'. The sub-band FDTD model fails to accurately predict such phenomenon due to the staircase approximation of the curved surfaces and such a problem can be alleviated by using a conformal FDTD method \cite{Hao}. Fig. \ref{2DPL} indicates that for modelling simple on-body communication scenarios such as both transmitter and receiver are on the trunk, UTD is very efficient and provides accurate results.

\subsection{Three-Dimensional On-Body Propagation Channels}
    Both the sub-band FDTD and UTD/RT are applied to model the UWB on-body radio channel in three dimensions. As shown in Fig. \ref{HumanBody} (a), different antenna positions are chosen due to locations of commonly used on-body communication devices such as head mounted display, headset, and wristwatch \textit{etc.} The human body is modelled by several different geometries: 1 sphere for the head ($r=0.10$ m), 1 ellipse cylinder for the trunk ($a=0.15$ m, $b=0.12$ m, $h=0.65$ m) and 4 cylinders for arms ($r=0.05$ m, $h=0.70$ m) and legs ($r=0.07$ m, $h=0.85$ m) according to measurement candidate's dimensions.

\begin{figure}[t]
    \centering
    \includegraphics[width=8.2cm,height=6cm]{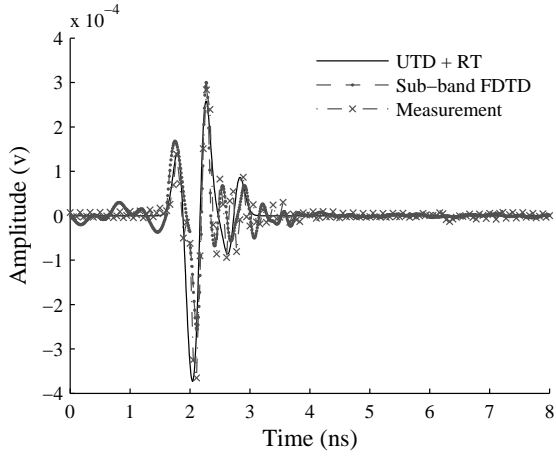}
    \caption{Comparison of channel impulse responses at Rx3 (Fig. \ref{HumanBody} (a)) using PICA from the UTD/RT model, sub-band FDTD model and measurement.}
    \label{Rx3PICA}
\end{figure}

\begin{figure}[t]
    \centering
    \includegraphics[width=8.2cm,height=6cm]{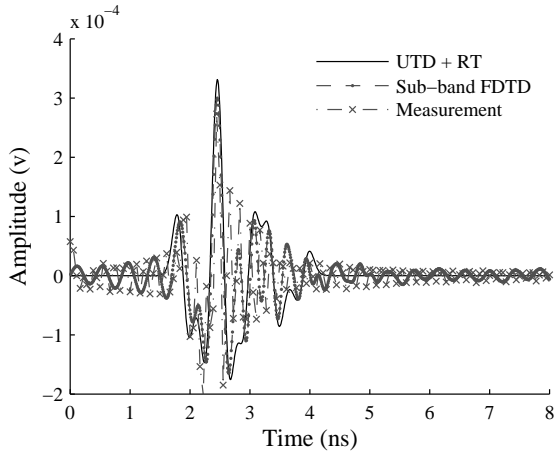}
    \caption{Comparison of channel impulse responses at Rx4 (Fig. \ref{HumanBody} (a)) using PICA from the UTD/RT model, sub-band FDTD model and measurement.}
    \label{Rx4PICA}
\end{figure}

    In the sub-band FDTD, the whole body model only consists of muscle with the dielectric constant and conductivity are obtained from measurement \cite{Tissues} (Fig. \ref{Permittivity}). The modelling environment is free space which is meshed by $140\times160\times630$ cells with each cell size $3.0\times10^{-3}$ m. The time step is chosen as $5.0\times10^{-12}$ s according to stability criterion. The PICA is modelled as a point source due to its omni-directional radiation properties. The channel impulse responses (CIRs) at two different receiver locations (Rx3 and Rx4 in Fig. \ref{HumanBody} (a)) are shown in Fig. \ref{Rx3PICA} and \ref{Rx4PICA}.

\begin{figure}[t]
    \centering
    \includegraphics[width=8.2cm,height=6cm]{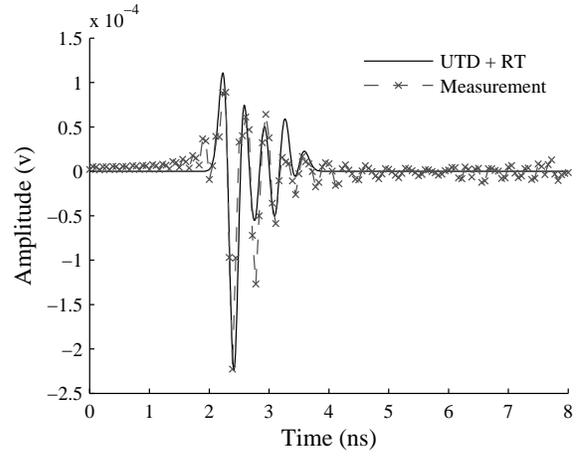}
    \caption{Comparison of channel impulse responses at Rx3 (Fig. \ref{HumanBody} (a)) using HSCA from the UTD/RT model and measurement.}
    \label{Rx3HSCA}
\end{figure}

\begin{figure}[t]
    \centering
    \includegraphics[width=8.2cm,height=6cm]{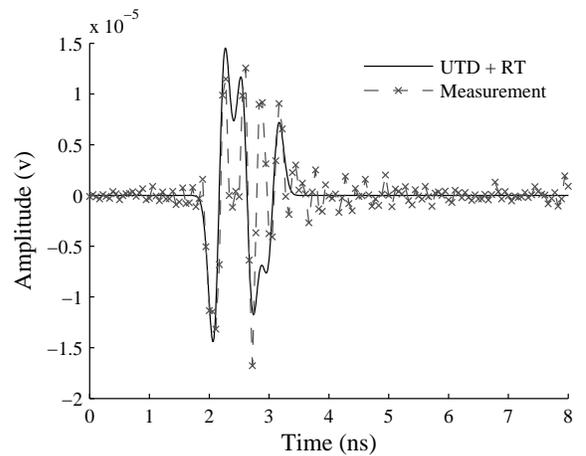}
    \caption{Comparison of channel impulse responses at Rx4 (Fig. \ref{HumanBody} (a)) using HSCA from the UTD/RT model and measurement.}
    \label{Rx4HSCA}
\end{figure}

    In UTD/RT, the ray tube angle is set to be $0.5^{\circ}$ for high accuracy \cite{Wang}. The tracing of rays and the calculation of diffracted field are performed following the procedure introduced in section III. Rays are terminated after their field strength drops 50 dB below the reference level. The threshold for on-body channel modelling is lower than that for indoor channel modelling (30 dB) due to the non-reflecting environment (free space) and relatively low amplitude of the received signal in our analysis. The CIRs at two different receiver locations (Rx3 and Rx4 in Fig. \ref{HumanBody} (a)) using two types of antennas (HSCA and PICA, section II) from the UTD/RT model are shown from Fig. \ref{Rx3PICA} to \ref{Rx4HSCA}.

    At the same receiver location (Rx3 or Rx4 in Fig. \ref{HumanBody} (a)), the CIR using PICA contains more multipath components compared with HSCA because of the difference between their radiation properties (Fig. \ref{Patterns}). Using the same antenna at different receiver locations, for PICA (Fig. \ref{Rx3PICA} and \ref{Rx4PICA}), sub-band FDTD provides more accurate results (in terms of the number of multipath components) than UTD/RT compared with measurement since FDTD can fully account for the effects of reflection, diffraction and radiation, while some rays are missing in UTD/RT model compared with measurement; for HSCA, greater difference has been observed between UTD/RT and measurement at Rx4 (Fig. \ref{Rx4HSCA}) compared with Rx3 (Fig. \ref{Rx3HSCA}), which is due to the more complicated scenario caused by severer body shadowing at Rx4 and higher order reflections/diffractions occur. For the sub-band FDTD model, the major difference between modelling results and measurements is caused by the approximation of antenna by a point source, and the change of antenna radiation patterns at different frequencies. While for the UTD/RT model, the difference is due to the approximation of human tissue as conducting material, and the change of antenna radiation patters at different frequencies.
\begin{figure}[t]
    \centering
    \includegraphics[width=8.2cm,height=6cm]{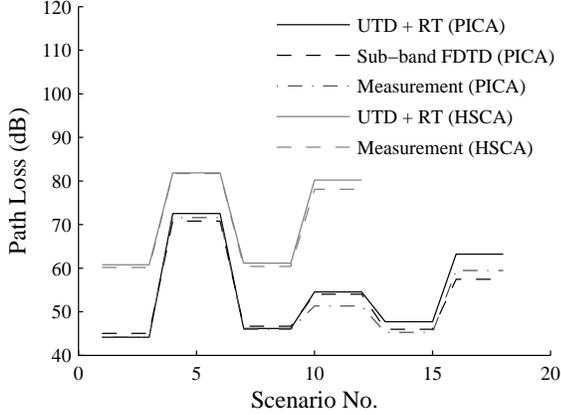}
    \caption{Comparison of the average path loss around each receiver location for different on-body scenarios using PICA (Rx1 - Rx6, Fig. \ref{HumanBody} (a)) and HSCA (Rx1 - Rx4), respectively.}
    \label{PLPICA}
\end{figure}

    In the local area of each receiver (Rx1 - Rx6 for PICA and Rx1 - Rx4 for HSCA, Fig. \ref{HumanBody} (a)), two more receiver locations are considered thus path loss values at total 18 different locations are obtained. Then the average path loss is calculated around each receiver location and compared with measurement results. Fig. \ref{PLPICA} shows the comparison of average path loss for different on-body scenarios using PICA and HSCA respectively. It can be seen that for the distinctive on-body radio channel, the path loss results from both sub-band FDTD and UTD/RT models are close to measurement.
\begin{figure}[t]
    \centering
    \includegraphics[width=8.2cm,height=6cm]{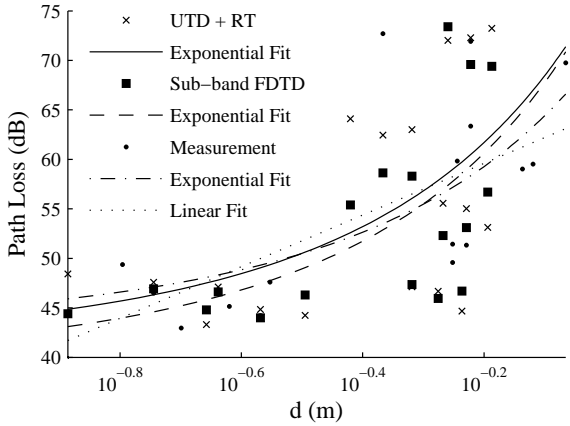}
    \caption{Comparison of path loss for UWB on-body channels using PICA from UTD/RT model, sub-band FDTD model and measurement. The exponential fitted curve for each model and the linear fitted line for measurement data are also shown.}
    \label{PLEPICA}
\end{figure}

    Different least-square (LS) fitting method can be used to fit the measurement and simulation data. In addition to the conventional linear power law fitting method, since the on-body channel mainly consists of creeping waves around the body, the exponential loss (dB per metre) fitting might also be appropriate. The exponential fitting can be performed using (\ref{eq_exponential}) \cite{Piazzi},
\begin{equation}
    PL(d)=Ae^{-\alpha d}
    \label{eq_exponential}
\end{equation}
where $A$ is the excitation coefficient and $\alpha$ is the attenuation coefficient. The linear power law fitting can be used to obtain the path loss exponent, $\gamma$ through (\ref{eq_ple}),
\begin{equation}
    PL(d)=PL(d_{0})+10\gamma\log_{10}(\frac{d}{d_{0}})+X_{\sigma}
    \label{eq_ple}
\end{equation}
where $PL(d_{0})$ is reference path loss at distance $d_{0}$, $\gamma$ is the path loss exponent, and $X_{\sigma}$ is a zero-mean Gaussian random variable.

\begin{figure}[t]
    \centering
    \includegraphics[width=8.2cm,height=6cm]{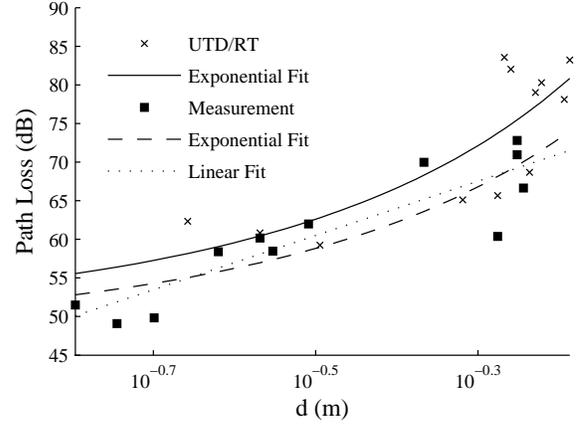}
    \caption{Comparison of path loss for UWB on-body channels using HSCA from UTD/RT model and measurement. The exponential fitted curves and the linear fitted line for measurement data are also shown.}
    \label{PLEHSCA}
\end{figure}

\begin{table}[t]
    \renewcommand{\arraystretch}{1.3}
    \caption{The values of excitation and attenuation coefficients from the UTD/RT, sub-band FDTD models and measurement using PICA.}
    \label{tbl_exponential_coefficients}
    \centering
    \begin{tabular}{l c c c}
        \hline
        & UTD+RT & Sub-band FDTD & Measurement\\
        \hline
        Excitation coefficient & 41.30 & 39.45 & 42.97\\
        \hline
        Attenuation coefficient & -0.64 & -0.68 & -0.51\\
        \hline
    \end{tabular}
\end{table}

\begin{table}[t]
    \renewcommand{\arraystretch}{1.3}
    \caption{The values of excitation and attenuation coefficients from the UTD/RT model and measurement using HSCA.}
    \label{tbl_exponential_coefficients2}
    \centering
    \begin{tabular}{l c c c}
        \hline
        & UTD+RT & Measurement\\
        \hline
        Excitation coefficient & 49.14 & 47.31\\
        \hline
        Attenuation coefficient & -0.77 & -0.69\\
        \hline
    \end{tabular}
\end{table}

    Fig. \ref{PLEPICA} and \ref{PLEHSCA} show the comparison of path loss for different on-body channels using PICA and HSCA respectively. The exponential fitted curves for different models and measurement are also shown. For comparison, the linear power law fitting is also performed for the measurement data. It can be seen that the exponential fitting is more appropriate for the data obtained due to the fact that the creeping wave is the dominant mechanism for the receiver locations considered in our models.

    The values of excitation and attenuation coefficients for UTD/RT, sub-band FDTD models and measurement for PICA and HSCA are listed in Table \ref{tbl_exponential_coefficients} and \ref{tbl_exponential_coefficients2}, respectively. For PICA, it can be seen that the UTD/RT model provides closer match to measurement compared with the sub-band FDTD. In the UTD/RT model, the measured antenna pattern is used and the human body is approximated by conducting sphere and cylinders. In the sub-band FDTD model, although the material frequency dispersion is considered, the antenna is approximated as a point source. The results indicate that the antenna pattern has more important effect than the material frequency dispersion on the on-body channel. For HSCA, the difference is mainly caused by the change of HSCA's radiation pattern at different frequencies and the approximation of human body as conducting structures in the UTD/RT model.

\section{Conclusion}
    The on-body radio channel modelling is performed using a sub-band FDTD and a combined UTD/RT model. In the sub-band FDTD model, the entire frequency band (3 - 9 GHz) is first divided into 12 sub-bands with 500 MHz bandwidth for each sub-band in order to take into account the material frequency dispersion at different frequencies. Within each sub-band, the conventional FDTD is applied to calculate the channel impulse responses. A combination technique is used at the receiver to recover all the sub-band simulations. The advantage of this method is its ability of modelling materials with any type of frequency dependence. In the UTD/RT model, the human body is approximated by conducting sphere and cylinders, and the ray tracing technique is used together with the generalised Fermat principle to find surface diffracted ray path while the UTD surface diffraction coefficients are used for calculating the received signal strength. The proposed models are applied to both 2-D and 3-D on-body scenarios and compared with measurement results. For cases such as both transmitter and receiver are mounted on the trunk, UTD/RT provides relatively simple and reliable solutions even if the human body is modelled as a conducting elliptic cylinder; while for more complicated scenario as for the whole body channel modelling, the sub-band FDTD is capable of providing more general solutions due to its ability of fully accounting for the effects of reflection and diffraction. The modelling results indicate that the antenna pattern has significant impacts on on-body radio channels. Through the UTD/RT approach, the effect of different antenna types on on-body radio channels is also investigated. The modelling results show good agreement with measurement.

\section*{Acknowledgment}
The authors would like to thank the reviewers for their valuable comments and suggestions. Thanks also to John Dupuy for assistance with measurements.

\begin{biography}[{\includegraphics[width=1in,height=1.25in,clip,keepaspectratio]{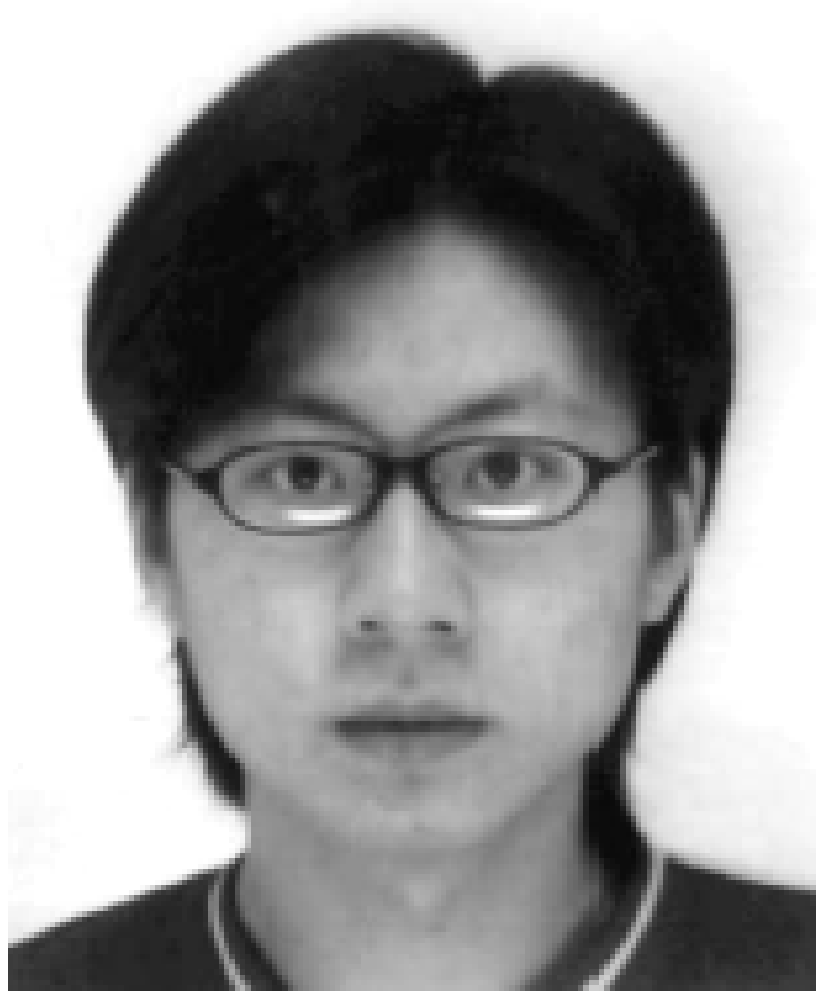}}]{Yan Zhao}
received the B.Sc. degree from Department of Information Engineering, Beijing University of Posts and Telecommunications (BUPT), China in 2002, and the M.Sc. degree from Department of Electronic, Electrical and Computer Engineering, The University of Birmingham, United Kingdom in 2003.

    Since 2003, he joined Department of Electronic Engineering, Queen Mary University of London, and currently working toward the Ph.D. degree. His main research interests include finite-difference time-domain (FDTD) modelling of dispersive materials, ultra-wideband (UWB) radio systems, and human interactions with indoor radio channels.
\end{biography}

\begin{biography}[{\includegraphics[width=1in,height=1.25in,clip,keepaspectratio]{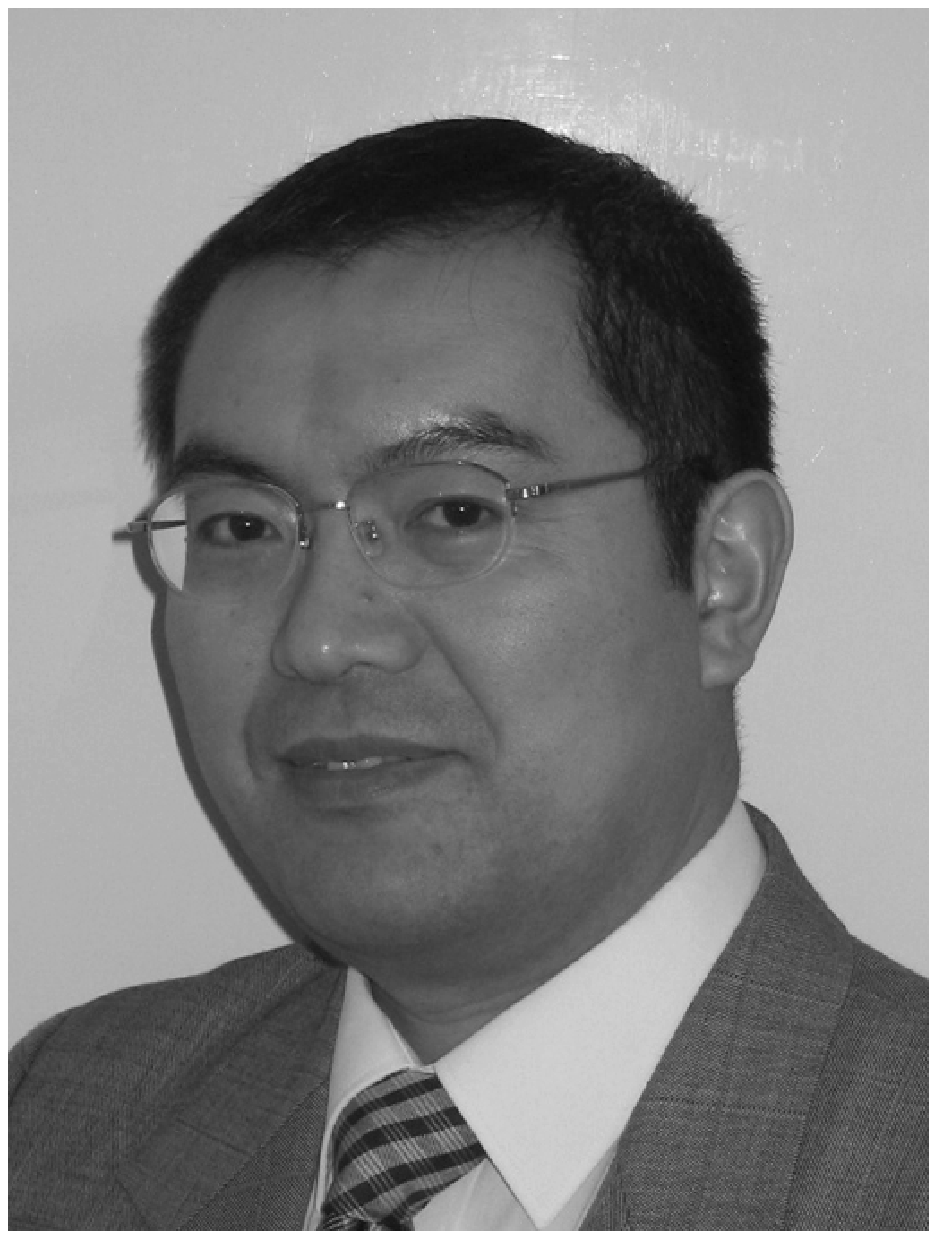}}]{Yang Hao}
(M'99) received the Ph.D. degree from the Centre for Communications Research (CCR) at the University of Bristol, U.K. in 1998. From 1998 to 2000, he was a postdoc research fellow at the School of Electrical and Electronic Engineering, University of Birmingham, U.K. In May 2000, he joined the Antenna Engineering Group, Queen Mary College, University of London, London, U.K. first as a lecturer and now a reader in antenna and electromagnetics. Dr. Hao has co-edited a book, contributed two book chapters and published over 60 technical papers. He was a session organiser and chair for various international conferences and also a keynote speaker at ANTEM 2005, France.

    Dr. Hao is a Member of IEE, UK. He is also a member of Technical Advisory Panel of IEE Antennas and Propagation Professional Network and a member of Wireless Onboard Spacecraft Working Group, ESTEC, ESA. His research interests are computational electromagnetics, on-body radio propagations, active integrated antennas, electromagnetic bandgap structures and microwave metamaterials.
\end{biography}

\begin{biography}[{\includegraphics[width=1in,height=1.25in,clip,keepaspectratio]{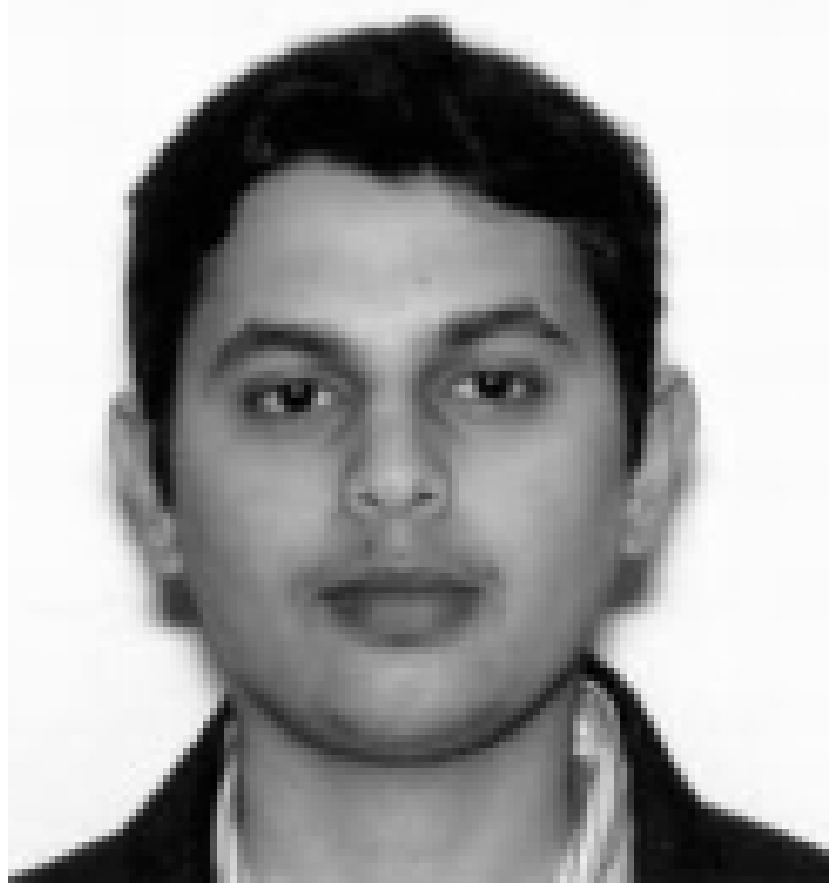}}]{Akram Alomainy}
received the M.Eng. degree in communication engineering from Queen Mary, University of London (QMUL), United Kingdom, in July, 2003.

    In September 2003, he commenced his research studies in the Electronic Engineering Department, QMUL, UK, where he is currently working toward the Ph.D. degree. His current research interests include, small and compact antennas for wireless body area networks, radio propagation characterisation and modelling for body-centric networks, antenna interactions with human body, computational electromagnetic and advanced antenna enhancement techniques.

    Mr. Alomainy has attended a number of established international conferences and workshops and has many referred conference and journal publications.
\end{biography}

\begin{biography}[{\includegraphics[width=1in,height=1.25in,clip,keepaspectratio]{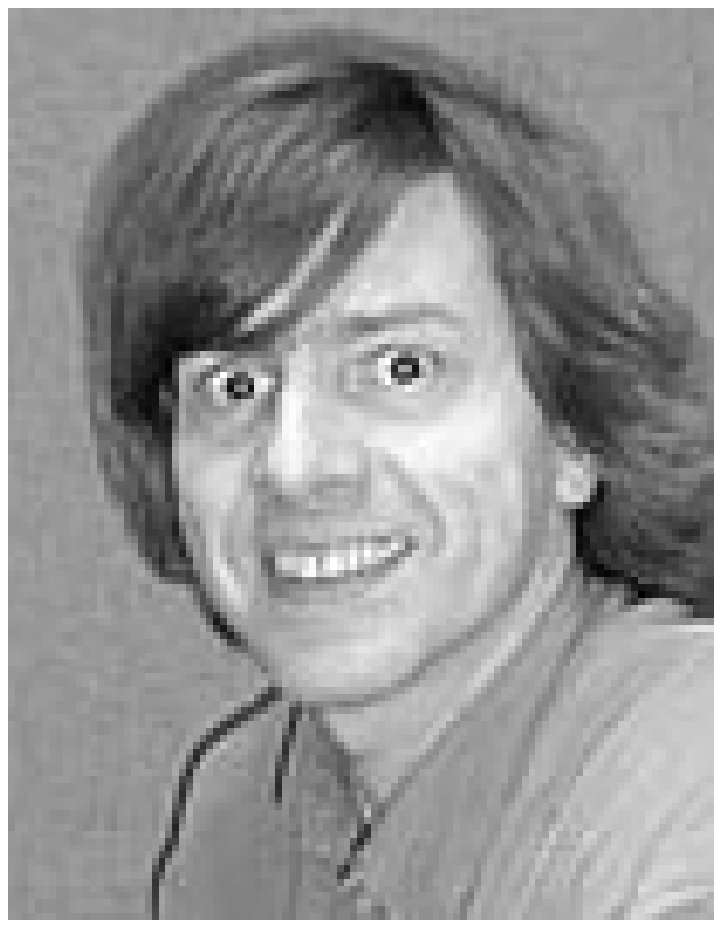}}]{Clive G. Parini}
(M'96) received the B.Sc.(Eng) and Ph.D. degrees from Queen Mary College, University of London, U.K., in 1973 and 1976, respectively.

    He then joined ERA Technology Ltd., U.K., working on the design of microwave feeds and offset reflector antennas. In 1977, he returned to Queen Mary College and is currently Professor of Antenna Engineering and Heads the Communications Research Group. He has published over 100 papers on research topics including array mutual coupling, array beam forming, antenna metrology, microstrip antennas, application of metamaterials, millimeterwave compact antenna test ranges, and millimeterwave integrated antennas.

    Prof. Parini is a Fellow of the Institution of Electrical Engineers (IEE), London, U.K. In January 1990, he was one of three co-workers to receive the Institution of Electrical Engineers (IEE) Measurements Prize for work on near field reflector metrology. He is currently the Chairman of the IEE Antennas and Propagation Professional Network Executive Team and is an Honorary Editor of \textit{IEE Proceedings Microwaves, Antennas and Propagation.} He has been on the organizing committee for a number of international conferences and in 1991 was the Vice Chairman and in 2001 the Chairman of the IEE International Conference on Antennas and Propagation.
\end{biography}

\balance

\end{document}